\documentstyle[own_sngl]{article}
\newcommand{\oversim}[2]{\protect{\mbox{\lower0.5ex\vbox{%
   \baselineskip=0pt\lineskip=0.2ex
   \ialign{$\mathsurround=0pt #1\hfil##\hfil$\crcr#2\crcr\sim\crcr}}}}} 
\newcommand{\simgreat}{\mbox{$~\mathrel{\mathpalette\oversim>}~$}} 
\begin{document}
%
\title {An Expanding Trapezium Cluster?\footnote{Poster presented at
the NATO ASI on The Physics of Star Formation and Early Stellar
Evolution (C.Lada \& N.Kylafis), Crete, 24.05.--5.06.98. Further
details and additional models will appear in Kroupa, Petr \&
McCaughrean (1998).}  }

\author {{\bf Pavel Kroupa}\\
\vskip 10mm
\small{Institut f{\"u}r Theoretische Astrophysik\\ 
Universit{\"a}t Heidelberg, Tiergartenstr. 15, D-69121 Heidelberg\\
e-mail: pavel@ita.uni-heidelberg.de\\
}} 

\begin{abstract}
\noindent 
Simulations with Aarseth's (1994) NBODY5 code are presented of an
initially dense binary-rich cluster. It is assumed that the star
formation efficiency is 50~per cent with instantaneous mass loss.

The model central density and velocity dispersion agree with the
observational constraints if expansion is only about $6\times 10^4$~yr
old. Additionally, the observed binary proportion constrains the
primordial proportion to have been significantly less than in
Taurus--Auriga.

The claim that a {\bf variation of the birth binary proportion with
gas cloud parameters has been detected}, however, can only be verified
if the cluster can be shown to be expanding rapidly.

\end{abstract}

\clearpage\newpage

\section{Introduction} 
\label{sec:intro}

\noindent
The Trapezium Cluster in Orion is an important astrophysical
laboratory for the study of the formation of stars and planetary
systems, and of the early dynamical evolution of a young cluster
rich in binary systems.  

\begin{itemize}
\item
It's {\underline {\bf age}} is estimated to be younger than 5~Myr, and most
probably only a few $10^5$~yr.
\item
The {\underline {\bf central stellar number density}} is unusually
large for embedded clusters, with mean inter-stellar separations of
about 6000~AU.
\end{itemize}

\noindent
The distribution of orbital elements of a population of {\underline
{\bf binary systems}} carries a {\bf memory of the dynamical history}
of this population.  Short-period primordial binary systems are an
{\bf energy reservoir}, whereas long-period binary systems are an {\bf
energy sink}.  The cluster and its constituent binary systems
interact, which in part {\bf drives cluster evolution}.

\begin{itemize}
\item
Surveys of {\bf Galactic field main-sequence stars} show that a
significant fraction ($f_{\rm tot}\approx\,50$\%) are in binary
systems (e.g. Duquennoy \& Mayor 1991).

\item
Observations of sparse groups of {\bf pre-main sequence stars in
Taurus-Auriga} imply a significant overabundance ($f_{\rm
tot}\approx1$) of binary systems (e.g. K\"ohler \& Leinert 1998).

\item
In the {\bf Trapezium Cluster} the binary proportion is comparable
to the Galactic field value (Petr et al. 1998a,b, Prosser et
al. 1994).
\end{itemize}

\noindent
{\underline {\bf QUESTION:}} Can a Taurus--Auriga binary proportion
have {\bf dynamically evolved} to the reduced value, or is the
Trapezium population {\bf primordial~?}

\clearpage\newpage

\section{Models}
\label{sec:models}

\subsection{The initial model Trapezium Cluster}
\label{sssec:cl}

\begin{itemize}
\item $N = 1600$ stars (point masses) (McCaughrean \& Stauffer 1994).
\item Half-mass radius $R_{0.5}=0.1$~pc (McCaughrean \& Stauffer 1994).
\item IMF from Kroupa et al.  (1993) for $m\le1\,M_\odot$, and Scalo
(1986) for $m>1M_\odot$.  
\item Lower and upper stellar mass limits of $m_{\rm
l}=0.08\,M_\odot$ and $m_{\rm u}=30\,M_\odot$, respectively.
\item Plummer density distribution.
\item Initial position and velocity vectors 
are independent of stellar mass.
\item Velocity distribution of the binary centre-of-masses is isotropic.
\end{itemize}

\vskip 5mm

\noindent
The resulting {\bf cluster mass} is $M_{\rm cl}=700\,M_\odot$.  
If the Trapezium Cluster were in virial equilibrium then the
model {\bf relaxation and crossing times} would be $t_{\rm relax}=0.62$~Myr and
$t_{\rm cross}=0.1$~Myr, respectively. 

\vskip 9mm

\noindent
The {\bf virial ratio} $Q=E_{\rm kin}/|E_{\rm pot}| = 0.5$ in virial
equilibrium. Here $Q=1$.

The initial velocities are chosen to correspond to a system with a
combined mass $M_{\rm stars}+M_{\rm gas}=2\times M_{\rm stars}$. That
is, it is assumed that the {\bf massive stars} in the cluster have
driven out a gas mass, $M_{\rm gas} = M_{\rm stars}=M_{\rm
cl}=700\,M_\odot$ (cf Churchwell 1997), immediately after they ``turn
on'' and before any significant stellar-dynamical processes occur.
This corresponds to a {\bf star-formation efficiency} of 50~per cent.

\clearpage\newpage

\subsection{Primordial binary systems}
\label{sssec:prbins}
\noindent
The total (summed over all periods) binary proportion is
\begin{equation}
f_{\rm tot} = {N_{\rm bin} \over N_{\rm bin}+N_{\rm sing}}, 
\label{eqn:binf}
\end{equation}
where $N_{\rm bin}$ and $N_{\rm sing}$ are the numbers of bound binary
and single star systems, respectively.  

\begin{itemize}
\item 
$f_{\rm tot}=1$ assumes that the binary-star {\bf properties do not
vary with star-forming conditions}, apart from the effects of
crowding, and that they are identical to what is observed in
Taurus--Auriga.
\item
$f_{\rm tot}=0.6$ assumes that the {\bf binary-star properties vary
with star-forming conditions} in the sense suggested by Durisen \&
Sterzik (1994).
\end{itemize}

\noindent
The {\bf initial model period distributions} are compared with the
observational data in {\underline{\bf Fig.~\ref{fig:init_P}}}. 

\vskip 5mm

\noindent
Additionally, the following assumptions are made:
\begin{itemize}
\item The initial {\it mass-ratio distribution} is obtained by random
pairing of the stars. 
\item The initial {\it eccentricity distribution} is thermally
relaxed.
\end{itemize}

\vskip 5mm

\noindent
{\underline {\bf Fig.~\ref{fig:e_p}}} shows the resulting {\bf initial
eccentricity-period} diagram, after pre-main sequence eigenevolution.

\clearpage\newpage

\section{Results}
\label{sec:res}
\noindent

\begin{itemize}
\item
The {\bf central density} and {\bf tangential velocity dispersion}
decrease rapidly due to the expansion. They are {\bf consistent with
the observed values} at time $t\approx0.06$~Myr.

\item
Rapid {\bf expansion halts the disruption of binary systems} at an early
stage.  
\end{itemize}

\vskip 5mm

This is shown in {\underline {\bf Figs.~\ref{fig:d_vdB} and~\ref{fig:fappB}}}.

\clearpage\newpage

\noindent
{\underline{\bf Implications:}
\begin{itemize}
\item
The expanding model could be a reasonable description of reality if
about 50~per cent of the mass of the cluster was expelled about
60~thousand years ago, at which time the present Trapezium Cluster
would have gone into a rapid expansion phase.  This solution implies
an {\bf extreme youth of the gas-free Trapezium Cluster}.

\vskip 3mm

This would be consistent with claims that the circum-stellar material
seen around the young stars in the Trapezium Cluster should be removed
within $10^4-10^5$~yr through {\bf photo-ionization} by the most
massive central star, $\theta^1$~C~Ori (e.g. Bally et al. 1998).

\item
The model apparent binary proportion is consistent with the
observational constraint for $t\simgreat0.05$~Myr, {\it provided}
$f_{\rm tot}\approx0.6$ initially!

\vskip 3mm

This is {\bf lower than the binary proportion in Taurus--Auriga}.  If
model~B2 does represent reality, then this may be due to dynamical
evolution in the proto-cluster prior to gas expulsion, or due to a
dependency on cloud temperature as suggested by Durisen \& Sterzik
(1994).
\end{itemize}

\vskip 3mm

\noindent
{\underline {\bf Main shortcoming:}} 

\noindent
In the present study the Trapezium Cluster is treated as an {\bf
isolated entity}.

In reality, it appears to be the core of the much more massive and
extended ONC, which is partially embedded in the parent elongated
molecular gas cloud (Hillenbrand \& Hartmann 1998).

\vskip 3mm

\clearpage\newpage

%

\clearpage

\begin{table}
{\small
\begin{minipage}[t]{20cm}
\hspace{3cm}
  \begin{tabular}{*{5}{c}{l}}
   \tableline\tableline
    Model  &$R_{0.5}$   &$Q$ &$f_{\rm tot}$ 
           &log$_{10}P_{\rm max}$ &period distribution \\
           &{\scriptsize (pc)} & & &(days)\\
    \tableline\tableline
    B1                 &0.10       &1.00 &1.0  &8.43 &K2\\
    B2                 &0.10       &1.00 &0.6  &11.0 &DM91\\
    \tableline\tableline
\end{tabular}
\end{minipage}
}
\caption{{\bf Initial conditions for the Trapezium Cluster
models}. Three simulations are performed for each model.  Column~5
contains the {\bf maximum binary-star period}.  The {\bf minimum
period} is $P_{\rm min}=1$~day in all cases. The form of the {\bf
period distribution} is defined in column~6, where K2 (Kroupa 1995a)
refers to a birth distribution that is consistent with young systems
in Taurus--Auriga, and DM91 (Duquennoy \& Mayor 1991) refers to the
Gaussian log-period distribution of the Galactic field. A $1\,M_\odot$
system has, with log$_{10}P_{\rm max}=8.43$ (11) a semi-major axis of
8200~AU ($4.2\times10^5$~AU).
\label{table:models}}
\end{table}
\clearpage
\newpage

\begin{figure}
\plotfiddle{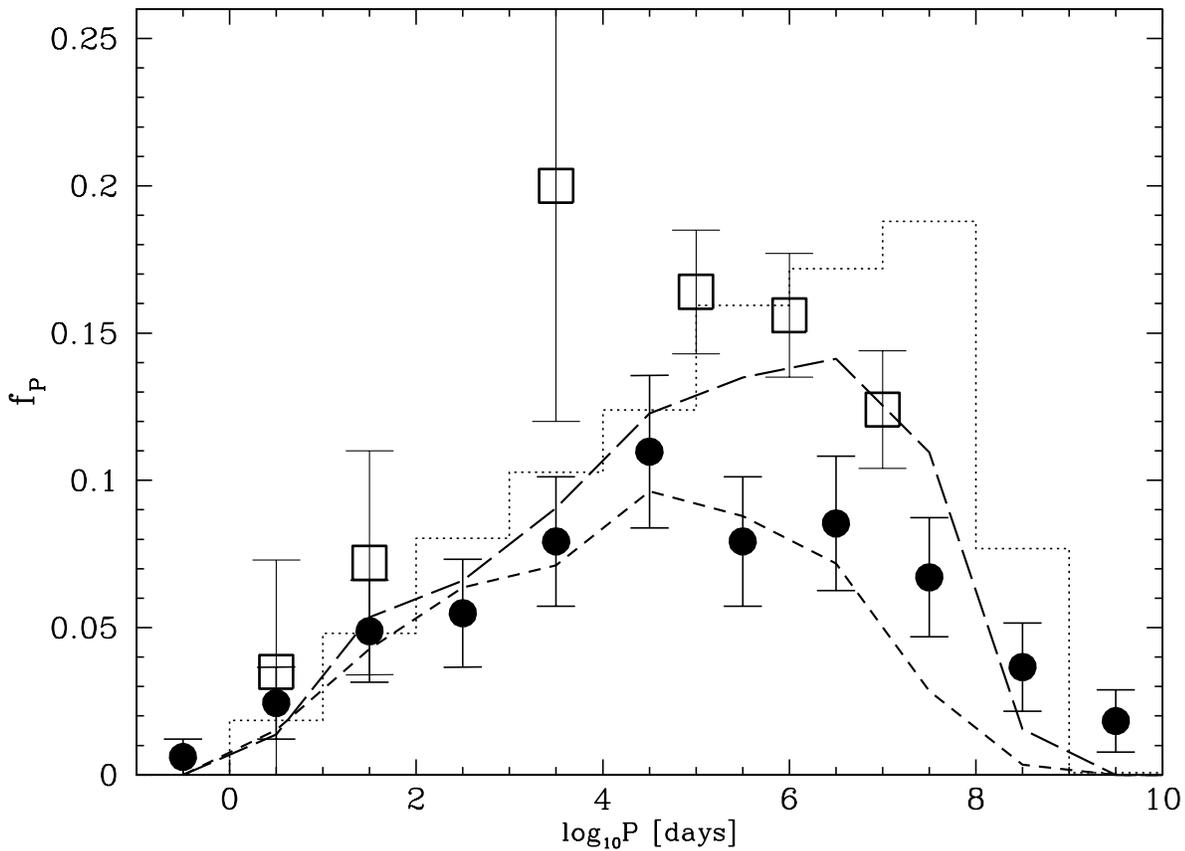}{15cm}{-90}{60}{60}{-250}{360}
\caption{{\bf Distribution of orbits}, $f_P$, for main sequence
multiple systems (solid dots, Duquennoy \& Mayor 1991) and pre-main
sequence systems in Taurus--Auriga (open squares; log$_{10}P>4$:
K\"ohler \& Leinert 1998, log$_{10}P=3.5$: Richichi et al. 1994,
log$_{10}P<2$: Mathieu 1994).  The dotted histogram is the initial
period distribution from Kroupa (1995a).  Crowding in the model
Trapezium Cluster changes this distribution to the long-dashed one. A
Gaussian log-period birth distribution that fits the solid dots,
changes in the model Trapezium Cluster due to crowding to the
distribution shown as the short-dashed line. 
\label{fig:init_P}}
\end{figure}

\clearpage
\newpage 

\begin{figure}
\plotfiddle{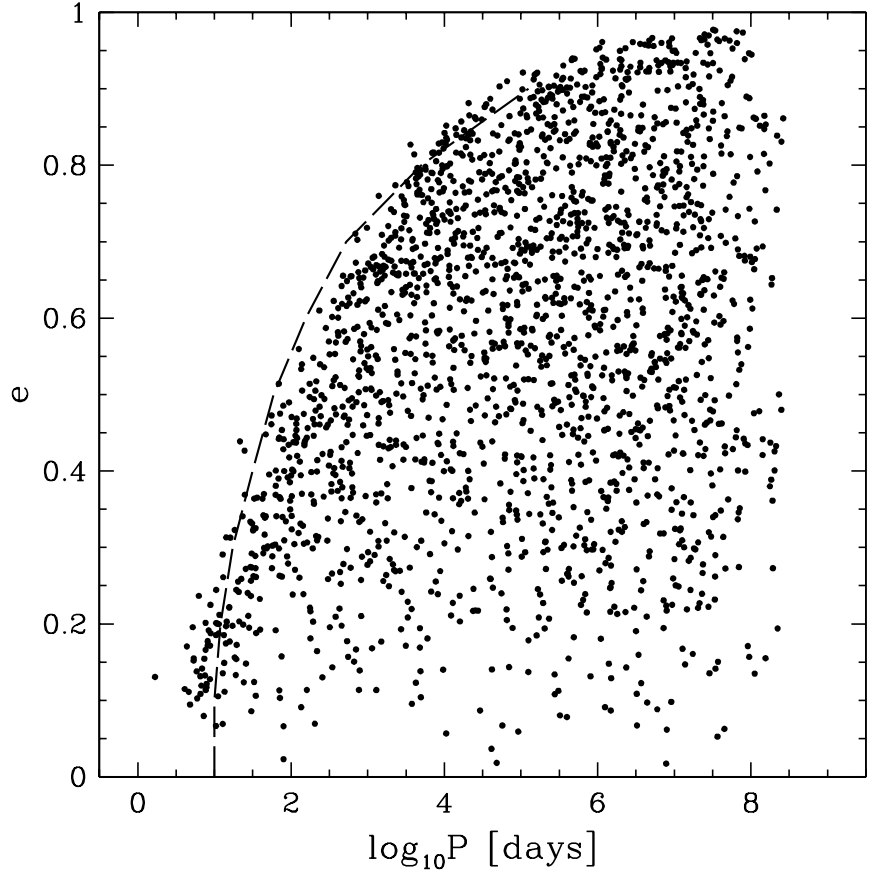}{12cm}{0}{60}{60}{-180}{-40}
\caption{The {\bf initial eccentricity--period diagram} for $f_{\rm
tot}=1$.  The distribution for $f_{\rm tot}=0.6$ is the same in
eccentricity, but has fewer orbits for log$_{10}P>4$. The thick dashed
line represents the observed envelope for main-sequence binary stars
with a G-dwarf primary (Duquennoy \& Mayor 1991).
\label{fig:e_p}}
\end{figure}

\clearpage\newpage

\begin{figure}
\plotfiddle{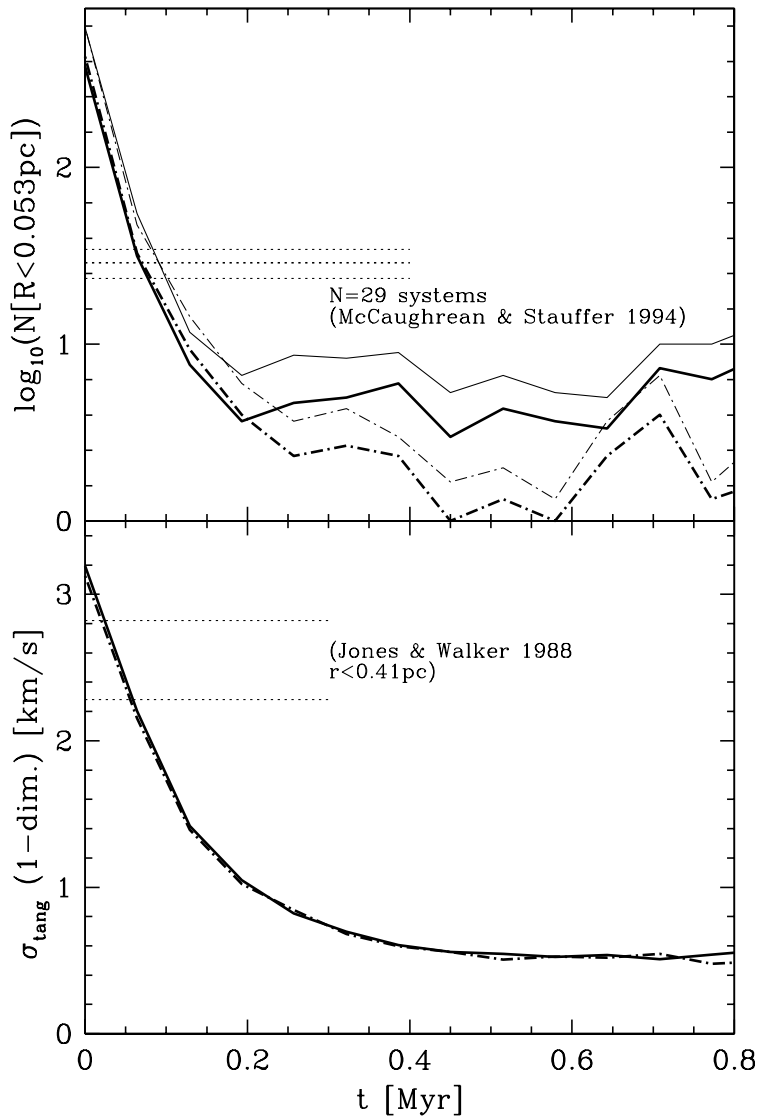}{15cm}{0}{60}{60}{-180}{-20}
\caption{ {\underline{\bf Upper panel:}} The {\bf time evolution of
the number of stellar systems} within a spherical distance of
$R=0.053$~pc of the position of the density maximum of the model
cluster.  The thick curves assume no binary systems are resolved, and
the thin curves count all stars.  {\underline{\bf Lower panel:}} The
{\bf time evolution of the velocity dispersion} of centre-of-masses
within a projected distance of $r=0.41$~pc of the position of the
density maximum of the model cluster.  {\underline{\bf In both
panels}}, the solid curves are for initially $f_{\rm tot}=1$
(model~B1), and the dot-dashed curves are for initially $f_{\rm
tot}=0.6$ (model~B2), and the observational constraints with the
Poisson error range are indicated by the dotted lines.
\label{fig:d_vdB}}
\end{figure}

\clearpage
\newpage 

\begin{figure}
\plotfiddle{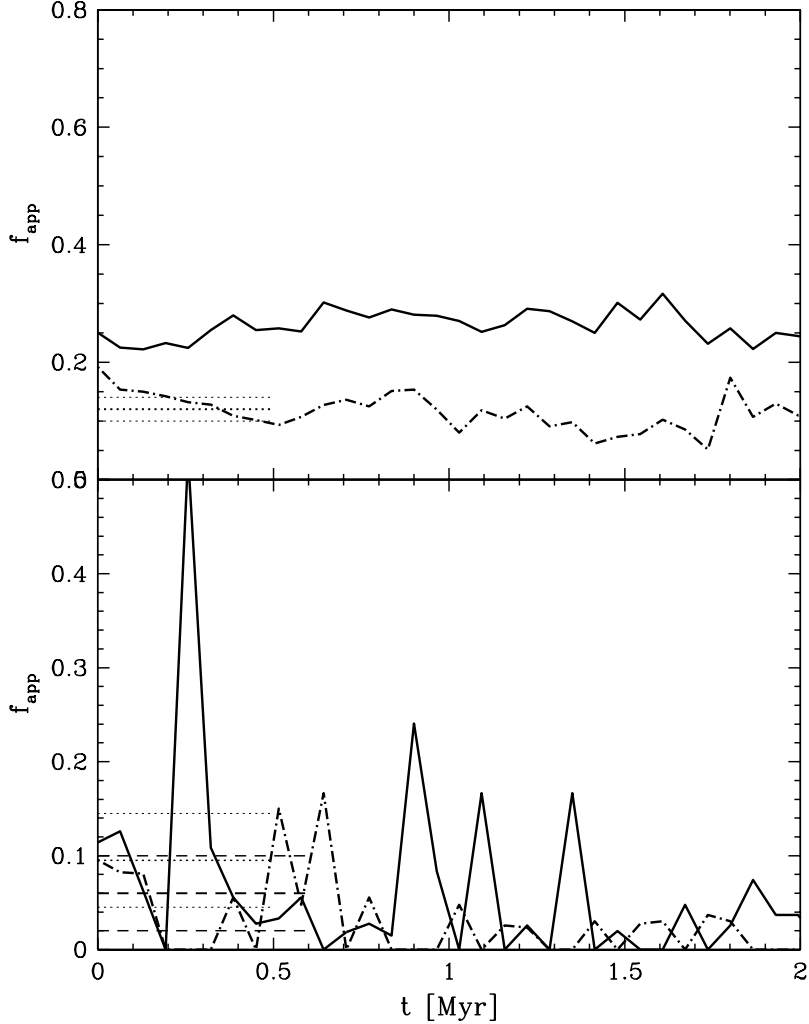}{14cm}{0}{55}{55}{-180}{-20}
\caption{ The {\bf time evolution of the apparent binary proportion}
in the model Trapezium Clusters.  The solid line is for initially
$f_{\rm tot}=1$ (model~B1) and the dot-dashed line is for initially
$f_{\rm tot}=0.6$ (model~B2). 
{\underline {\bf Upper panel:}} observational constraints
from Prosser et al. (1994, $r=0.249$~pc, $d_1=26$~AU, $d_2=440$~AU)
are shown as dotted lines.  
{\underline {\bf Lower panel:}} observational constraints
from Petr et al. (1998a, $r=0.04$~pc, $d_1=63$~AU, $d_2=225$~AU) are
shown as the horizontal lines. Here, the central dotted line is
$f_{\rm app}$ for all systems in their sample, and the central dashed
line is $f_{\rm app}$ for the low-mass systems only. Poisson
uncertainties are indicated by the upper and lower horizontal lines.
\label{fig:fappB}}
\end{figure}

\end{document}